\documentclass[]{spie}  %>>> use for US letter paper

\usepackage{amsmath,amsfonts,amssymb}
\usepackage[pdftex]{graphicx}
\usepackage[colorlinks=true, allcolors=blue]{hyperref}
\usepackage{wrapfig} 
\title{The DESI Instrument Control System: Status and Early Testing}

\author[a]{K.~Honscheid}
\author[a]{A.~E.~Elliott}
\author[b]{E.~Buckley-Geer}
\author[f]{B.~Abreshi}
\author[c]{F.~Castander}
\author[d]{L.~daCosta}
\author[b]{S.~Kent}
\author[e]{D.~Kirkby}
\author[f]{R.~Marshall}
\author[b]{E.~Neilsen}
\author[d]{R.~Ogando}
\author[g]{D.~Rabinowitz}
\author[h]{A.~Roodman}
\author[c]{S.~Serrano}
\author[i]{D.~Brooks}
\author[j]{M.~Levi}
\author[k]{G.~Tarle}

\affil[a]{The Ohio State University, Columbus, OH, USA}
\affil[b]{Fermi National Accelerator Laboratory, Batavia, IL, USA}
\affil[c]{Institut d'Estudis Espacials de Catalunya, Barcelona, Spain}
\affil[d]{Observatorio National, Rio de Janeiro, Brazil}
\affil[e]{University of California Irvine, Irvine, CA, USA}
\affil[f]{National Optical Astronomy Observatory, Tucson, AZ, USA}
\affil[g]{Yale University, New Haven, CT, USA}
\affil[h]{SLAC National Accelerator Laboratory, Menlo Park, CA, USA}
\affil[i]{Department of Physics \& Astronomy, University College London, London, UK}
\affil[j]{Lawrence Berkeley National Laboratory, Berkeley, CA, USA}
\affil[k]{Physics Department, University of Michigan, Ann Arbor, MI, USA}

\authorinfo{Corresponding author kh@physics.osu.edu}
\usepackage{import}      % cjb  - allow graphics files in same directory at tex made easy
\usepackage{xspace}		% manage spaces after newcommands, eg \lya

\newcommand{\simlt}{\lower.5ex\hbox{$\; \buildrel < \over \sim \;$}}
\newcommand{\simgt}{\lower.5ex\hbox{$\; \buildrel > \over \sim \;$}}

\begin{document}
\maketitle
%\linenumbers
\begin{abstract}
The Dark Energy Spectroscopic Instrument (DESI) is a new instrument currently under construction for the Mayall 4-m telescope at Kitt Peak National Observatory. It will consist of a wide-field optical corrector with a 3.2 degree diameter field of view, a focal plane with 5,000 robotically controlled fiber positioners and 10 fiber-fed broad-band  spectrographs. The DESI Instrument Control System (ICS) coordinates fiber positioner operations, interfaces to the Mayall telescope control system, monitors operating conditions, reads out the 30 spectrograph CCDs and provides observer support and data quality monitoring.
In this article, we summarize the ICS design, review the current status of the project and present results from a multi-stage test plan that was developed to ensure the system is fully operational by the time the instrument arrives at the observatory in 2019.
 
\end{abstract}

\section{The Dark Energy Spectroscopic Instrument and Survey}

The DESI collaboration is planning a five-year spectroscopic survey
covering 14,000 deg$^2$ of the sky, targeting three classes of galaxies during dark time
identified from imaging data. More than 30 million
\begin{figure}[h]
\centering
\includegraphics[width=6in]{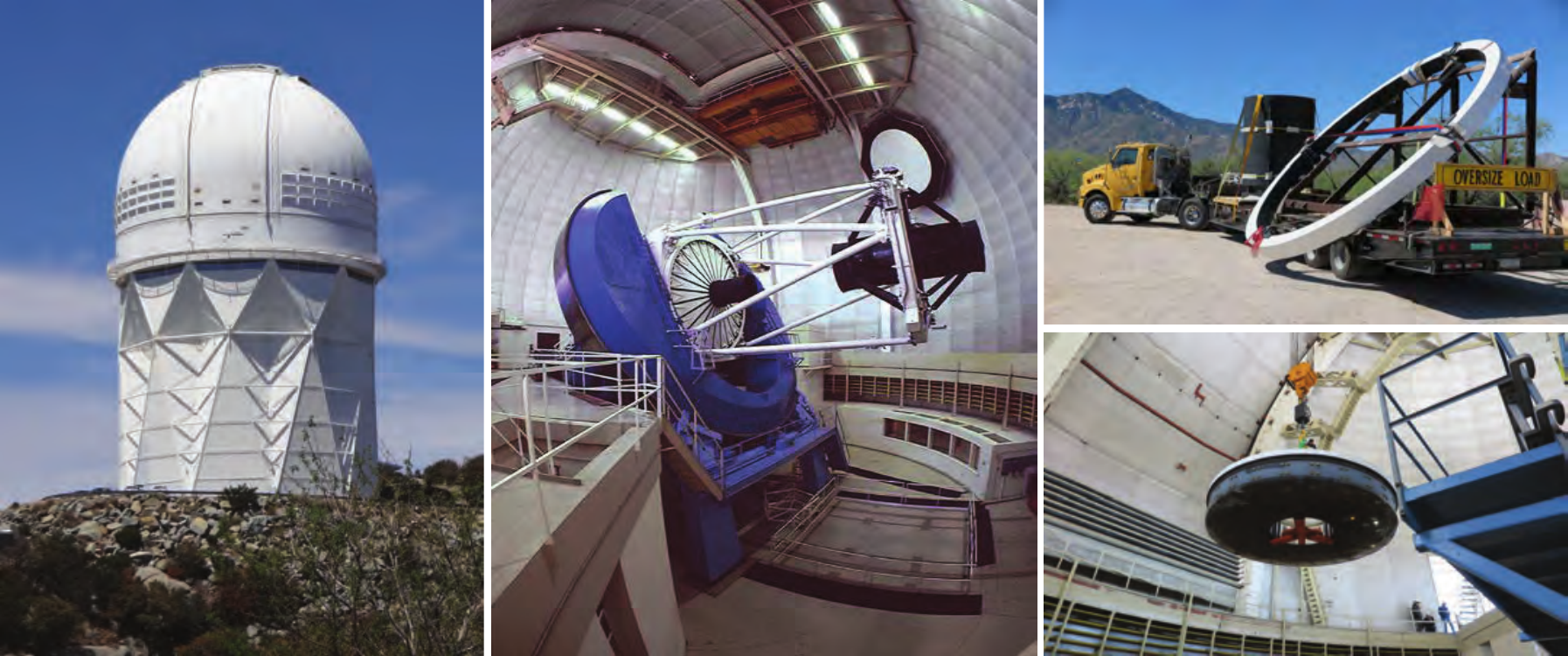}
\caption{Shown on the left is the Mayall 4m telescope dome.  The center image shows the telescope.  The black cylinder is the MOSAIC corrector that will be replaced as part of the DESI project to instrument a 3 degree diameter field of view. The photos on the right show the new DESI prime focus cage and top ring arriving at the observatory. The Mayall primary mirror has already been removed to prepare for DESI installation}
\label{fig:Mayall}
\end{figure}
galaxy and quasar redshifts will be obtained to measure the imprint of baryon acoustic oscillations (BAO)
and determine the matter power spectrum, including redshift
space distortions.  These measurements offer new insights into the accelerating expansion of the universe by constraining both the cosmological expansion history and the growth of large-scale structure. DESI will also provide cutting-edge limits on the sum of neutrino masses and provide new measurements to constrain theories of inflation.

DESI builds on the successful Stage-III BOSS redshift
survey, which has established BAO as a precision technique for dark
energy exploration.  DESI will make an order-of-magnitude advance over
BOSS in volume observed and galaxy redshifts measured 
by using 5,000 robotically controlled fiber-positioners to feed a
collection of spectrographs covering the wavelength range 360 nm to 980 nm.  Newly designed
optics for the National Optical Astronomy Observatory's 4-m Mayall telescope at Kitt Peak, Arizona, will
provide an 8-square-degree field of view.  The telescope dome and telescope with the existing MOSAIC corrector are shown in Figure~\ref{fig:Mayall}. DESI will be the first Stage IV dark energy experiment
and complements the imaging surveys Dark Energy Survey (DES, operating
2013-2018) and the Large Synoptic Survey Telescope (LSST, planned
start early in the next decade).

\section{The Instrument Control System}

The DESI instrument control system (ICS) is tasked to
perform all control and monitor functions required for successful operation of the DESI instrument.
This includes orchestration of the inter-exposure sequence to position both the telescope and the 5,000 fiber actuators for the next exposure, readout of the 30 CCDs in the ten spectrographs and the observer console.
The design of the DESI online system is based on the readout and control system architecture developed for the Dark Energy Camera \cite{Diehl2015}. This system was deployed on Cerro Tololo in 2012 and has been used successfully for both the DES survey and the community observing program. A detailed description of the DECam data acquisition system can be found in \cite{SPIEhonscheid2012} and \cite{SPIEhonscheid2014}. 

Requirements on the instrument control system flow directly from the DESI science goals and survey design. In order to maximize survey efficiency, the ICS has to reposition the 5,000 actuators onto new targets and complete all activities between exposures in less than 120 seconds with a design goal to achieve this in 60 seconds. A lower limit on the time between exposures is set by the CCD readout time of 42 seconds. The ICS design is required to be robust and modular in the sense that the system shall meet operating requirements even with some components failing so that system up-times of greater than 97\% can be achieved by the end of the first year of the 
survey. The ICS has to continuously monitor the operational parameters of the DESI instrument and must archive these parameters and other telemetry information for offline analysis. Observer support has to include consoles with graphical user interfaces, alarm and error notification systems and support for secure remote access.  An additional important requirement on the ICS architecture is the need to support small-scale test systems and sub-component integration tasks.
\begin{figure}[htb]
\centering
\includegraphics[width=0.7\textwidth]{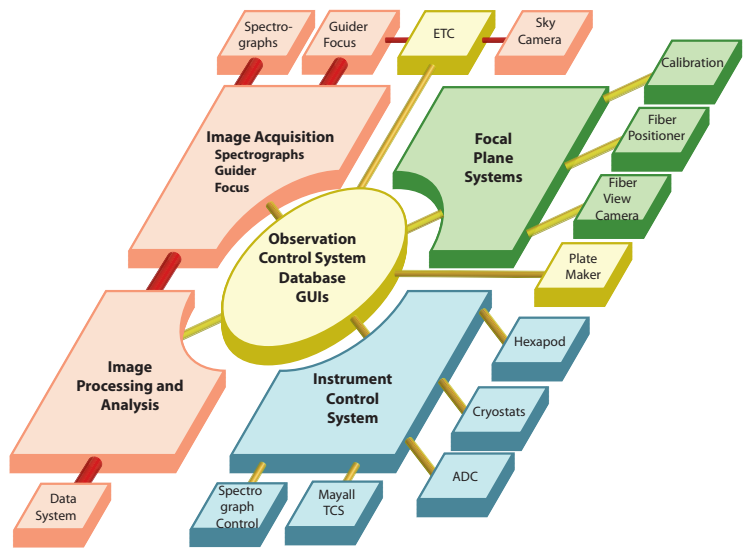}
\caption{Schematic view of the DESI readout and control system. The observer console and other user interfaces are not shown.}
\label{fig:daqoverview}
\end{figure}

The design of the data acquisition system we developed for DESI reflects these requirements. The overall architecture is shown schematically in Figure~\ref{fig:daqoverview}. Three key functions of the ICS can be distinguished: instrument control and monitoring, data readout and handling, and observation control.

 The ICS monitors every component of the instrument  and detailed information about instrument status, operating conditions and performance are archived in the DESI operations database. DOS, the software component of the DESI ICS provides all components with easy to use access to the telemetry and alarm history databases even with the rest of the system is offline.  The connection to the Mayall telescope control system (TCS) and the Mayall telemetry database is provided by a dedicated TCS Interface component also shown in Figure~\ref{fig:daqoverview}.
 \begin{figure}[htb]
\centering
\includegraphics[width=0.7\textwidth]{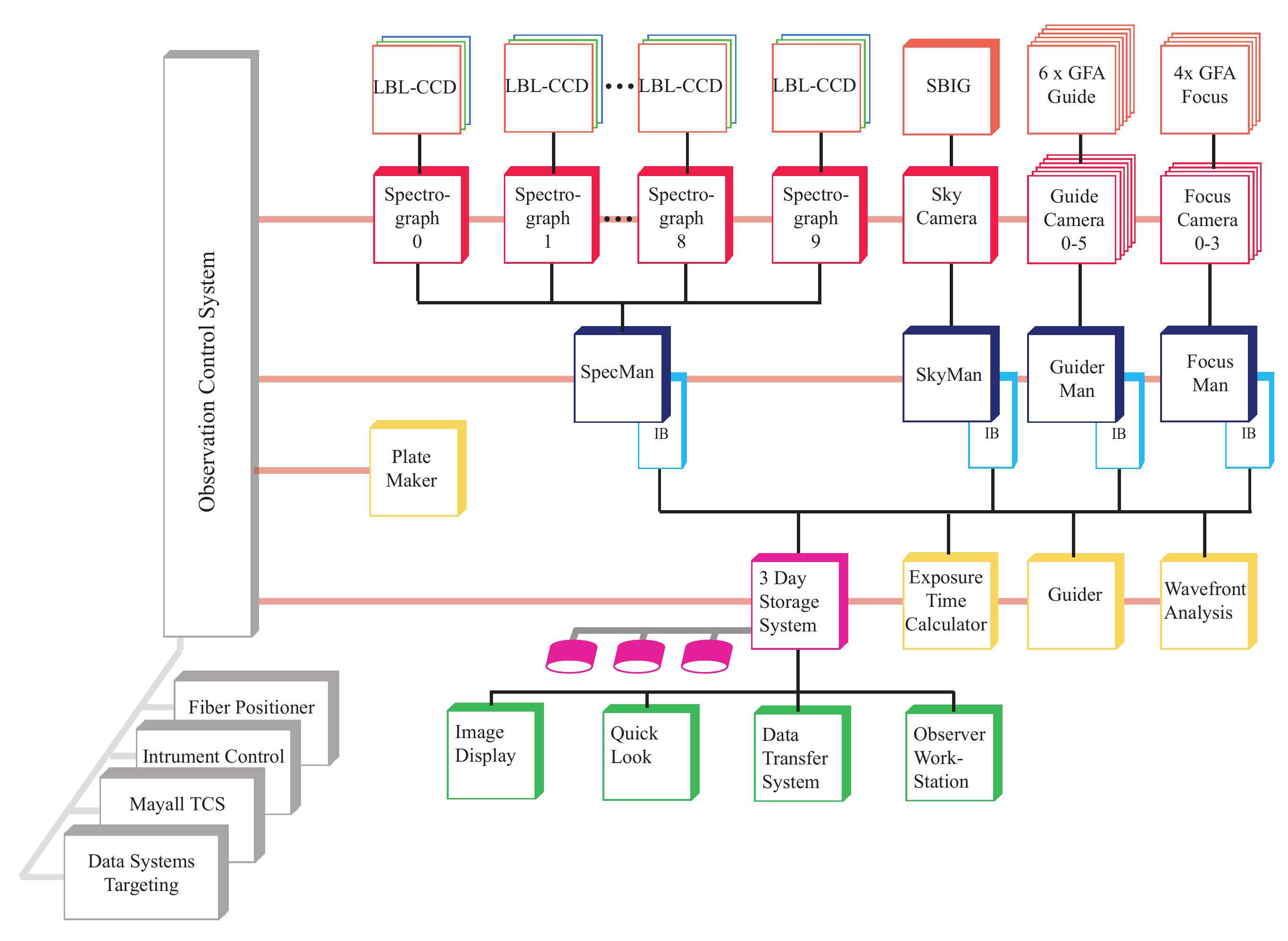}
\caption{Block diagram of the DESI data acquisition system. Data sources are shown in the top row and data flows from top to bottom. The observation control system that coordinates all activities is shown to the left. Data is stored in multi-extension FITS files on the 3-day storage disk array and is distributed to the data transfer system \cite{SPIEfitzpatrick2010}, quality assurance (QuickLook) and the observer workstations.}
\label{fig:DESIDAQ}
\end{figure} 
 At the core of the ICS sits the observation control system or OCS that orchestrates the complex DESI inter-exposure sequence. The OCS interacts closely with the focal plane systems including the fiber positioner, the fiber view camera, and the guider, focus and alignment system. Data is acquired from the guider, focus and an auxiliary camera to measure sky levels as well as the CCD front end electronics in the spectrographs. Data handling is parallelized to maximize throughput until the data are saved to disk.
From here the images are picked up by the DESI Data Systems group and transferred off the mountain to the processing center at LBNL. Data are also distributed to online data quality assurance tools and the observer workstations for interactive analysis.

The fourth and final component of the DESI online system - not shown in Figure~\ref{fig:daqoverview} are the observer consoles with web-based user interfaces that include the instrument and survey control, alarm history as well as telemetry and status displays. 
\begin{wrapfigure}{R}{0.45\textwidth}     %%%%[htb]
\centering
\includegraphics[width=0.4\textwidth]{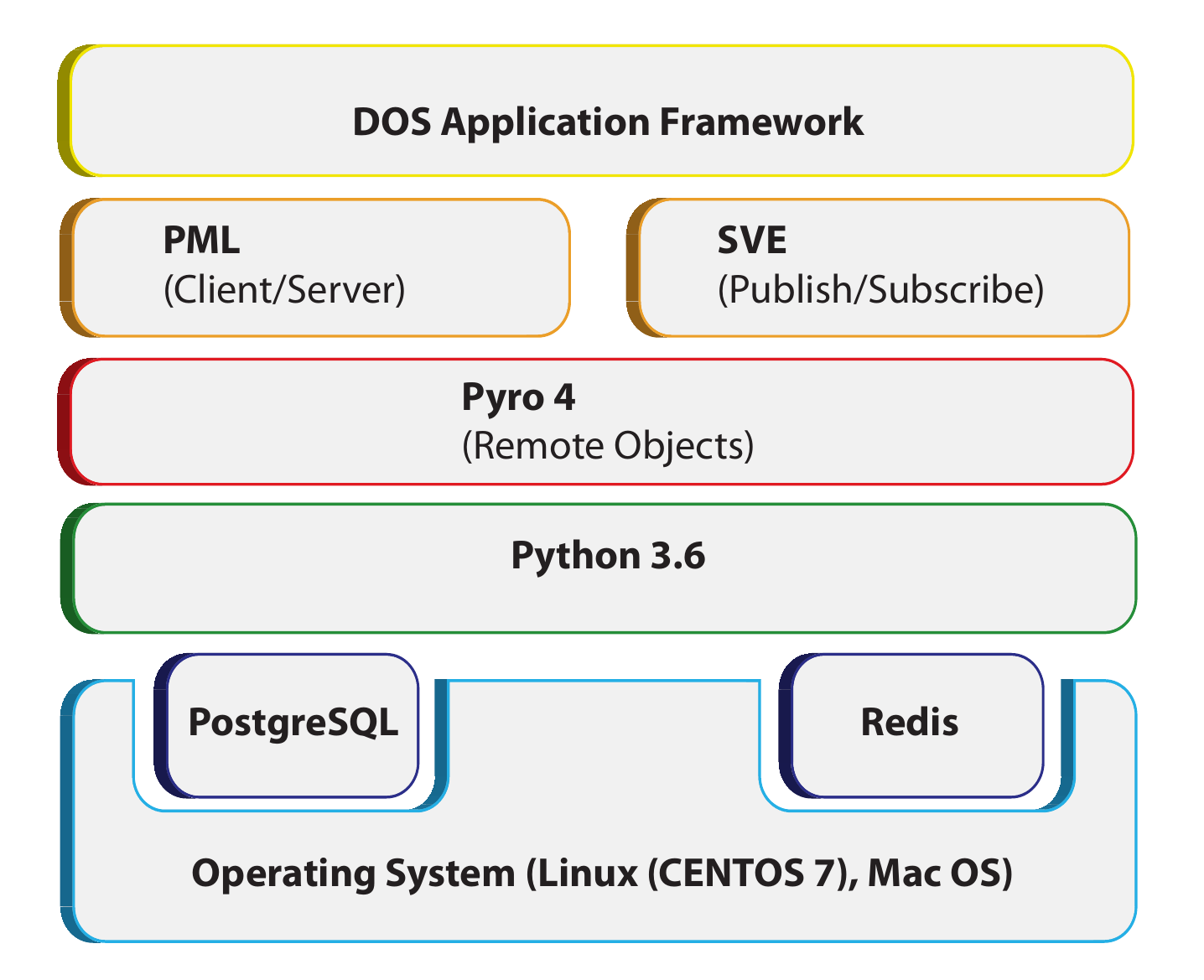}
\caption{ The DOS software stack: All DESI online computers run Centos 7. Application software is written in Python 3.6. System services such as logging, alarms, message passing and database access are provided via a standardized application framework.}
\label{fig:dossoftware}
\end{wrapfigure}

 The DESI online system is implemented as a distributed multi-processor system. All together it consists of more than 100 applications running on about 70 nodes ranging from embedded device controllers to server-class workstations. A more hardware oriented view of the system is show in Figure~\ref{fig:DESIDAQ}. Every node is connected to the network - the device controllers use Gigabit Ethernet whereas the rackmount computers operate at 10 gigabit speed.

The 5,000 fibers in the DESI focal plane feed ten identical spectrographs each with three cameras covering different 
wavelength regions. The spectra produced  by each camera are recorded by 4k$\times$4k CCDs with four readout amplifiers operating in parallel. 
A default pixel clock of 100~kpixels/s results in a readout time of approximately 42 seconds. The charge 
contained in each pixel is reported as a 16-bit unsigned integer yielding a data volume of 34~MBytes per camera or about 1~GByte per exposure for all 10 spectrographs. The images are compressed using the standard Rice/tile algorithm and saved to the 3-Day storage system. This disk array has sufficient capacity to record  several nights worth of images should the link off the mountain be unavailable for an extended period.

With the exception of the custom-designed spectrograph CCD controllers and the controllers for the focal plane cameras (guide, focus) all ICS hardware and network components are commercial, off-the-shelf products. The same philosophy was applied to the ICS software (DOS). DOS runs on Linux and MacOS operating systems and is entirely based on standard, open-source software. The DESI computers run Centos 7. The software is almost entirely written in Python, currently using Python 3.6. The operations database is based on PostgreSQL and we use Redis for the implementation of our shared variable (publish/subscribe) system. A schematic view of the DOS software stack is given in Figure~\ref{fig:dossoftware}. 

Given the ICS's distributed architecture, inter-process communication takes a central place in the design of the DESI infrastructure software. Our middleware solution is based on the Python Remote Objects (Pyro4) software package. Pyro4 is a free, advanced and powerful distributed object technology system written entirely in Python. It allows objects to interact just like normal Python objects even when they are spread over different computers on the network. Pyro4 handles the network communication transparently. Pyro4 provides an object-oriented form of remote procedure calls similar to Java's Remote Method Invocation (RMI). A name server supports dynamic object location, which renders (network) configuration files obsolete. Using the name server, DESI processes can locate and establish communication links with each other irrespective of the underlying hardware architecture; an important feature for any multi-processor communication system. For DESI we added an additional protocol based on the well known zeroconf/bonjour mechanism that  extends this concept to allow embedded controllers and small devices to operate even with the main DOS instance not running.
Using the services provided by Pyro4 and taking advantage of the high performance of the Redis in-memory database, we added customized client/server and publish/subscribe protocols to the DOS software stack. On top of that sits a standardized application framework that provides access to system services such as the telemetry database, logging, alarm messages and inter-process communication. Development of the DOS infrastructure software is complete allowing DOS to be used for the DESI prototype tests we describe in the next sections.  A detailed description of the DESI instrument control system can be found in \cite{SPIEhonscheid2016}.

\section{Functional verification and Testing of the DESI ICS}
\label{sec:tests}

The DESI survey will begin with a science validation phase in late 2019. The ICS, however, has to be ready a year earlier to support functional acceptance testing of DESI components arriving at the observatory and on-sky commissioning starting a year from now. 
\begin{figure}[htb]
\centering
\includegraphics[width=0.7\textwidth]{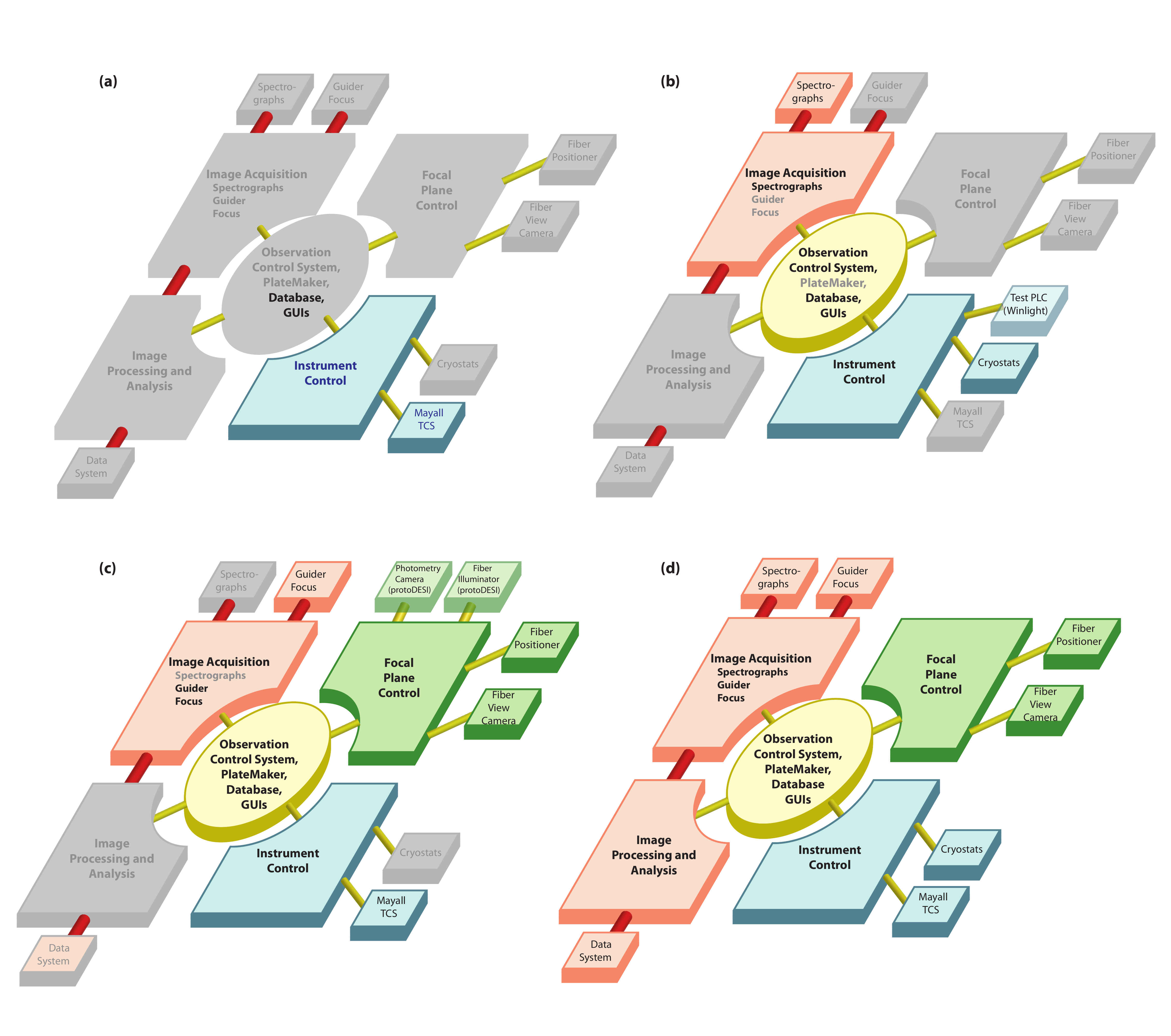}
\caption{ICS development cycles: (a) Mayall TCS integration (b) Spectrograph test stand at Winlight (c) ProtoDESI system (d) Full system test of the DESI ICS using hardware emulators.}. 
\label{fig:daqdevelopment}
\end{figure}
A suite of increasingly complex tests was developed and executed to ensure that the ICS is ready and operational in time. The four major tests are shown in Figure~\ref{fig:daqdevelopment} using the ICS architecture diagram from Figure~\ref{fig:daqoverview} to indicate which components participated. Details on each test follow in the next sections. We are currently at stage 4 performing full system tests using the complete ICS computer system that was recently installed in the telescope dome.

\subsection{Test 1: Mayall Telescope Interface}
\label{sec:TCSInterface}

The DESI Online System connects with the existing Mayall telescope control system (TCS) \cite{SPIEsprayberry2016} to communicate new pointing coordinates and to send correction signals derived from the guider. 
\begin{figure}[htb]
\centering
\includegraphics[width=0.8\textwidth]{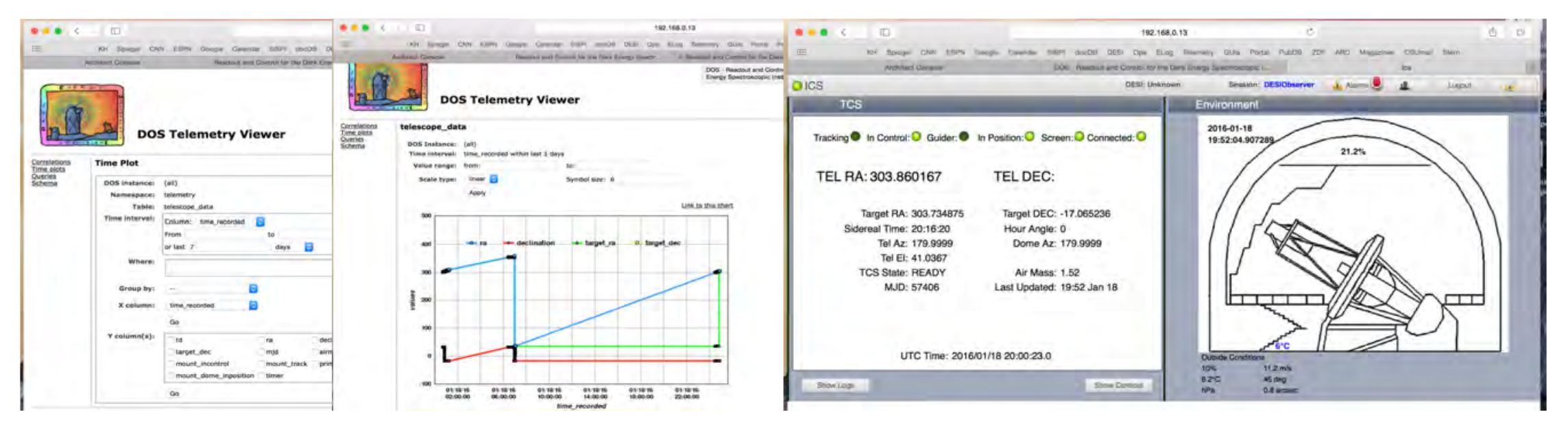}
\caption{Screenshots taken during the ICS-Mayall TCS test showing some of the DOS GUIs and the result of slew commands sent from the ICS to the telescope.}
\label{fig:tcstest}
\end{figure}
In return DESI receives telescope position and status information from the TCS. 
Commands are initiated by the ICS and communicated to the TCS via a socket connection running a custom protocol. Status and telemetry information including weather and dome environmental data are available to the ICS via a Redis database operated by the observatory.

The {TCS Interface} application and the connection between DOS and the Mayall TCS were tested during a 2 day/night test run at KPNO in January 2016. We successfully moved the telescope under DOS control, read back dome, telescope and environmental data and observed the telescope react to guider correction signals generated on the DOS side by a guider emulator. Screenshots taken during this test (Figure~\ref{fig:tcstest}) show changes in telescope position in response to DOS commands sent to the TCS. In addition to the TCS Interface this exercise provided a first test of the DOS infrastructure software and the web-based user interface system.
Updates to the Mayall TCS  environment software updates were completed in Spring 2017 and the ICS integration of the Mayall Redis database was done shortly thereafter. 

\subsection{Test 2: Spectrographs}
\label{sec:specttest}

The DESI spectrographs are built by Winlight, an optics company in southern France, and fully tested on-site by DESI collaborators from France.
For this purpose, a teststand with full spectrograph control was needed providing us with an opportunity to test the ICS software for CCD readout, image building, control of the spectrograph mechanisms and the interface to the independent CCD cryostat control system. The software was complete by the end of summer 2016 and spectrograph tests commenced in September of that year.
\begin{figure}[!htb]
\centering
\includegraphics[width=0.9\textwidth]{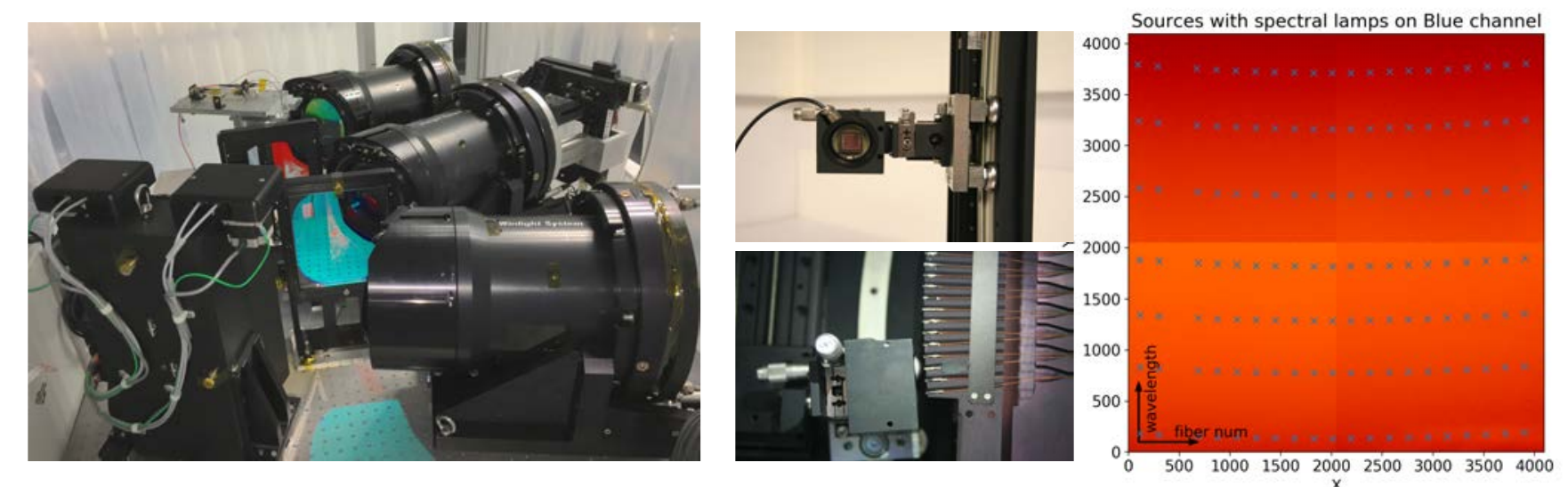}
\caption{Images from the DESI Spectrograph teststand at Winlight, France. The three spectrograph cameras are seen on the left. The center figure shows some of the test equipment to illuminate individual fibers and an image from the blue camera/CCD with spectra from a test slit illuminated by calibration lamps overlaid.}
\label{fig:spectrograph}
\end{figure}
The spectrograph test stand has been in routine operation since 2016 providing valuable confirmation of the reliability of the DOS infrastructure software and validating the CCD readout and spectrograph control software for use with the full DESI instrument. 
DESI online applications are based on a common software model. This DOS Application Framework  serves as a base class and gives the same basic structure to all applications. It provides a unified interface to all DESI  services such as the configuration system, the DOS data cloud (Shared Variables), the message passing and remote procedure system, alarms, logging, the constants database and the interlock and monitoring system. The Framework manages all resources centrally which allows it to unsubscribe all shared variables and to close all communication channels when the application is about to exit. Proper exit handling is critical for the ability to stop and restart individual processes without the need to take down the entire online system. Additional functionality provided by the Application Framework include a heartbeat that can be used to monitor the overall state of the system and process control functions.
The DOS Application Framework significantly reduces the complexity of application development and simplifies integration with other DOS components. Software developed by our French collaborators for the test equipment such as the moveable fiber illuminator shown in Figure~\ref{fig:spectrograph} uses the Application Framework.

\subsection{Test 3: ProtoDESI}
\label{sec:protodesi}

\begin{figure}[htb]
\centering
\includegraphics[width=0.9\textwidth]{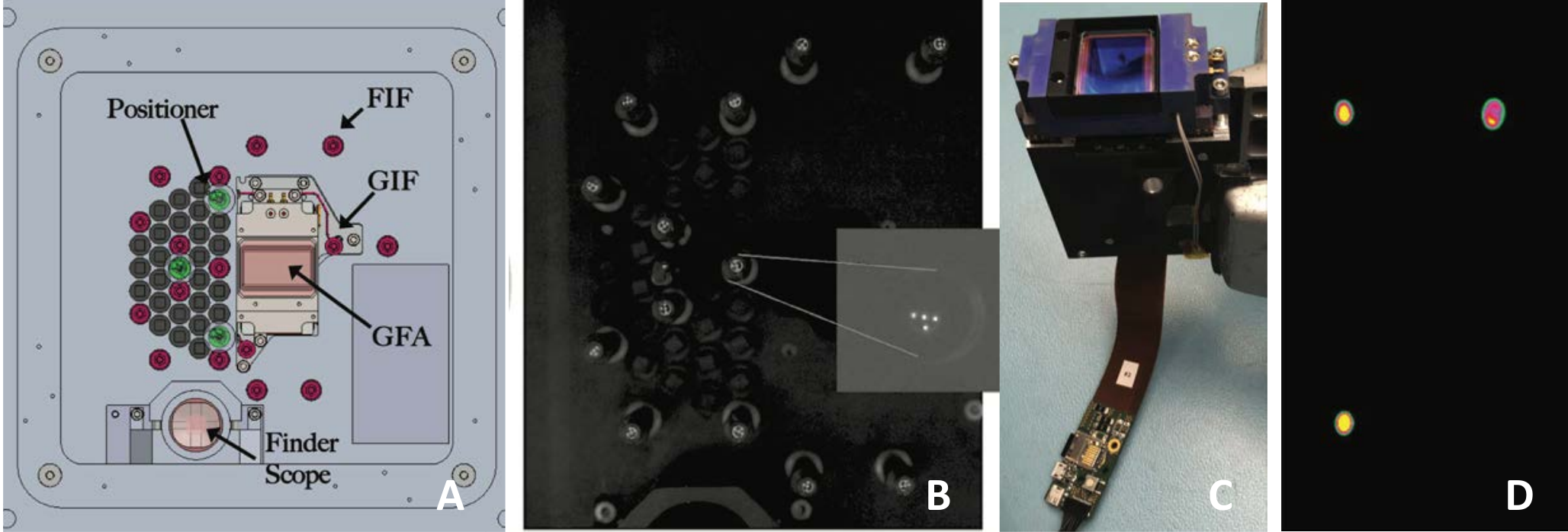}
\caption{The ProtoDESI instrument: (A) schematic view of the focal plane showing positioners, illuminated fiducials and the guide camera (GFA) (B) a photo of the ProtoDESI focal plane with the fiducials illuminated. Each fiducial has four spots. (C) the Guide Focus Assembly (GFA) - the ProtoDESI guide camera. (D) Glorious Success: three positioners on target illuminated by star light.}
\label{fig:protodesi}
\end{figure}

The ProtoDESI test in Summer 2016 was designed to validate fiber positioning and to demonstrate that we can successfully capture light from on-sky targets. Figure~\ref{fig:protodesi} shows a schematic view of the ProtoDESI focal plane as well as photos of several components. The success of the ProtoDESI test run is evidenced by the rightmost image showing three positioners on target illuminated by star light. A detailed description of ProtoDESI can be found in \cite{protodesi2018}.

The fiber positioning loop is orchestrated by the OCS. At the start of a new exposure the telescope is instructed to slew to the new field. The DESI hexapod and ADC are adjusted at the same time and the positioners perform an open loop move to the requested target coordinates. Once the telescope, ADC and hexapod are in position, the focus and alignment cameras  take a full-frame exposure. The out-of-focus star images (donuts) are analyzed to determine the wavefront and small adjustments are sent to the hexapod to complete telescope and focal plane alignment.
\begin{figure}[htb]
\centering
\includegraphics[width=0.85\textwidth]{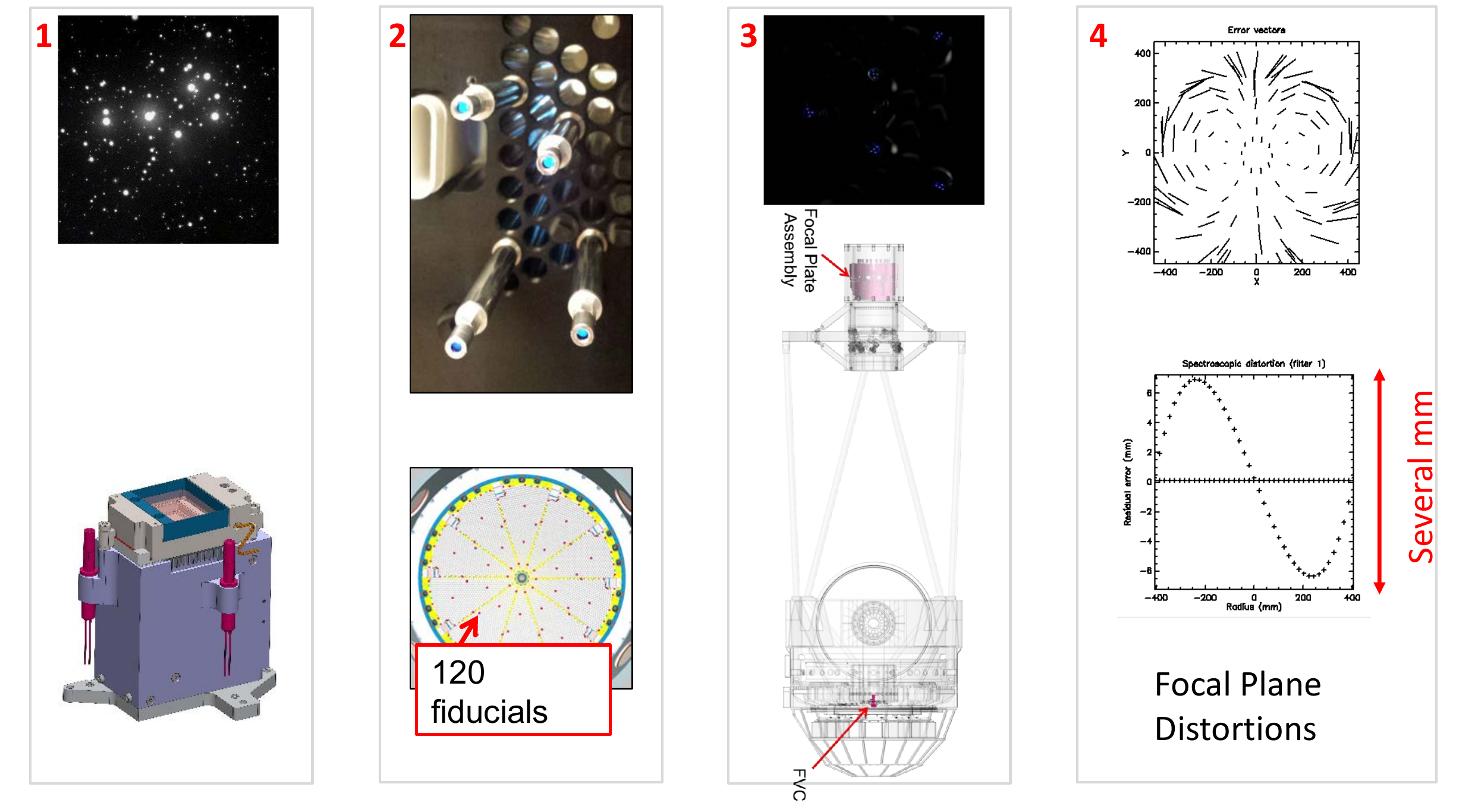}
\caption{Components pf the positioning loop: (1) Full frame GFA images are used to find the on-sky position (astrometry) (2) The positioners and fiducials are illuminated and imaged (3) by the fiber view camera mounted in the Cassegrain cage. (4) Measured fiber positions are mapped to sky positions by the PlateMaker software using a model of the optical distortions. 2-3 iterations are needed to position the actuators on target.}
\label{fig:positionerloop}
\end{figure}

The OCS then takes (in-focus) images with the guide cameras and instructs the PlateMaker application to find the astrometric solution. If necessary, the telescope pointing is corrected to center the field on the requested coordinates. Focal plane fiducials are illuminated and fibers are backlit to allow the fiber view camera to take an image of the current positions. The PlateMaker has the complex task to convert fiber view camera pixels to on sky as well as focal plane coordinates. Offsets between actual and requested positions are determined for each positioner and corrections are send to the focal plane. The last steps are repeated until the positioners are within 10 microns of the target position.
\begin{figure}[htb]
\centering
\includegraphics[width=0.9\textwidth]{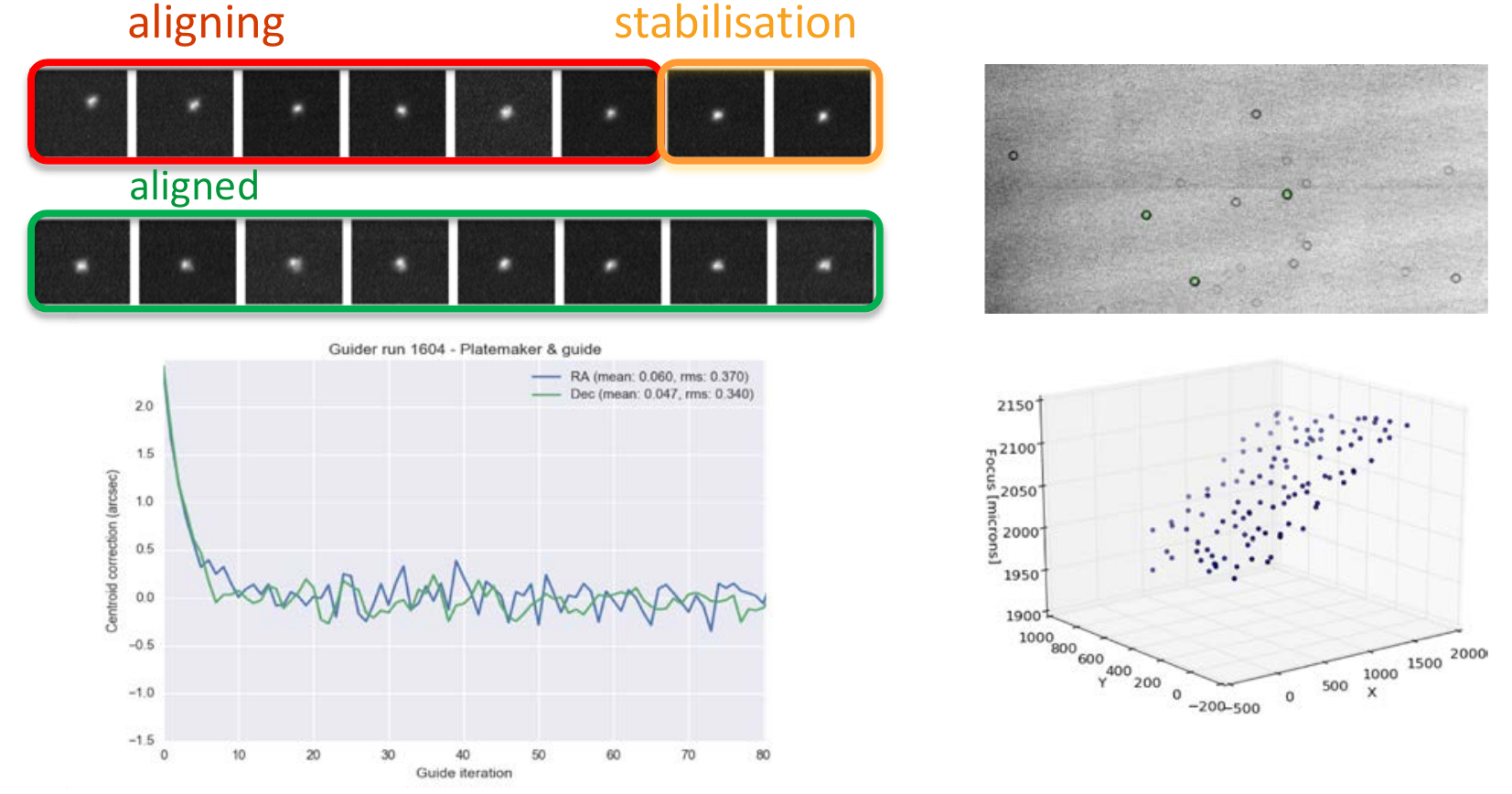}
\caption{ProtoDESI results: On the left, a sequence of guide star images showing the alignment phase where guide corrections to the TCS are used to move the star image to the requested pixel. The plot below shows the size of the corrections sent to the TCS during alignment and tracking phases. On the right, an out-of-focus image of star (donuts) and the focal plane tilt as measured by the Active Optics software (AOS)}
\label{fig:protodesiresults}
\end{figure}

For the ICS, ProtoDESI provided validation for a number of components: the exposure sequencer in the OCS, guide camera readout, fiber view camera software and operation and most importantly the PlateMaker concept that provides the critical coordinate transformations between sky and focal plane coordinates. ProtoDESI required a fairly large number of applications and processes which provided a good test of DOS configuration system. Not directly ICS software but quite important for DESI operations, we were able to optimize the layout of the new DESI control room in the Mayall dome  and evaluate hardware components for the observer consoles and the compute nodes.

Another major component of the ProtoDESI test was evaluation of the GFA guide camera and the guider software.
The guider's task is to analyze the images taken by the GFA detectors,
to determine the position of the guide star centroids and to derive
corrections signals that are sent to the TCS. Tracking uncertainties
have to be less than 0.03~arcsec. Multiple guide stars will be combined to achieve the required performance. Studies showed that near the galactic pole on average more than 10 sufficiently bright stars will be available. The DESI guider distinguishes two stages: the acquisition stage where guiding stars are selected and aligned to the request (pixel) location and the tracking stage where guide stars are continuously monitored and corrections are sent to the telescope.
By default, the DESI guider operates in catalog mode, where guide stars have been preselected from an external catalog. Also supported is self mode, where the guider analyzes the entire full-frame GFA images, and automatically selects the best guide stars, based on signal-to-noise, isolation, saturation,  and location with respect to the chip boundaries. Accurate positioning of the telescope is critical for DESI. The list of guidestars is included in the exposure request together with the fiber positioner targets. Once the PlateMaker has found the astrometric solution it converts the guidestar coordinates to pixel positions which the OCS sends to the guider. Typically, the guidestar will be close but not exactly on the right pixel. The guider determines the offset and sends correction signals to the TCS. During this alignment phase the guide star centroid shifts to the target position as shown in Figure~\ref{fig:protodesiresults}.

ProtoDESI did not have a focus and alignment camera nor was the instrument mounted with a hexapod, but we could deliberately move the instrument out of focus and record donut data for analysis by the active optic component (AOS) of the ICS \cite{SPIEroodman2014}. Results of these studies are shown on the right in Figure~\ref{fig:protodesiresults}. The top image shows a typical out-of-focus donut image. The AOS code was adapted to the MOSAIC corrector optics that was used for ProtoDESI and wavefront reconstruction showed a small tilt in the focal plane alignment that was confirmed when the instrument was shipped back to LBL by metrology measurements.

\subsection{Test 4: Full System Test and Mock Observing}
\label{sec:MockObs}

\begin{figure}[htb]
\centering
\includegraphics[width=0.8\textwidth]{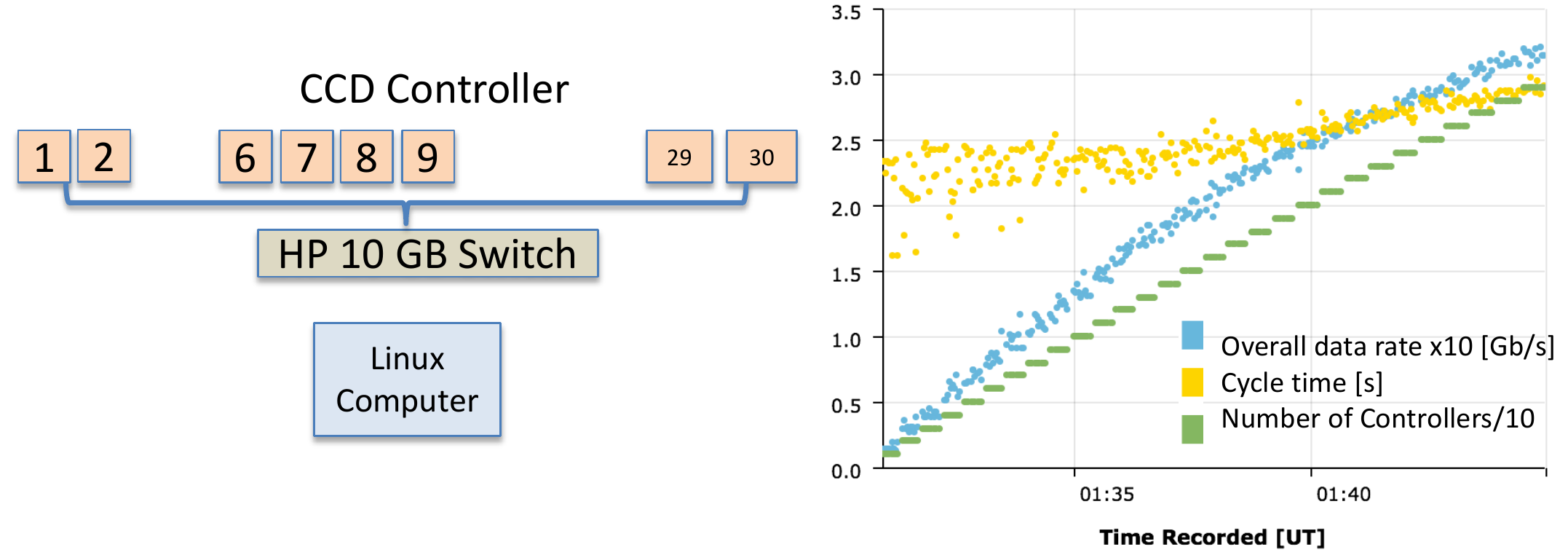}
\caption{Full system test using 30 CCD controllers. As the number of CCD controller increases (green), the overall throughput (blue) increases but the time per cycle stays roughly the same (yellow) indicating sufficient network and switch capacity and that the overall rate is set by the readout speed of the CCD controllers. Collecting data from all CCDs in 2-3 seconds is more than adequate for DESI. }
\label{fig:networktest}
\end{figure}

Ongoing full system tests are used to demonstrate that the ICS and in particular the infrastructure software scales to full DESI without any network  throughput or performance issues. A test simulating the expected data flow (Figure~\ref{fig:networktest}) confirmed more than adequate performance for DESI. Tests to full DESI scale were also performed for the databases. One of the more demanding uses of the DESI mountain top data base is recording information on each of the 5,000 positioners for every move. Using a full-scale simulation of the positioner system we were able to confirm that our schema design is sufficiently fast to handle the expected number of database operations.

\begin{figure}[htb]
\centering
\includegraphics[width=0.8\textwidth]{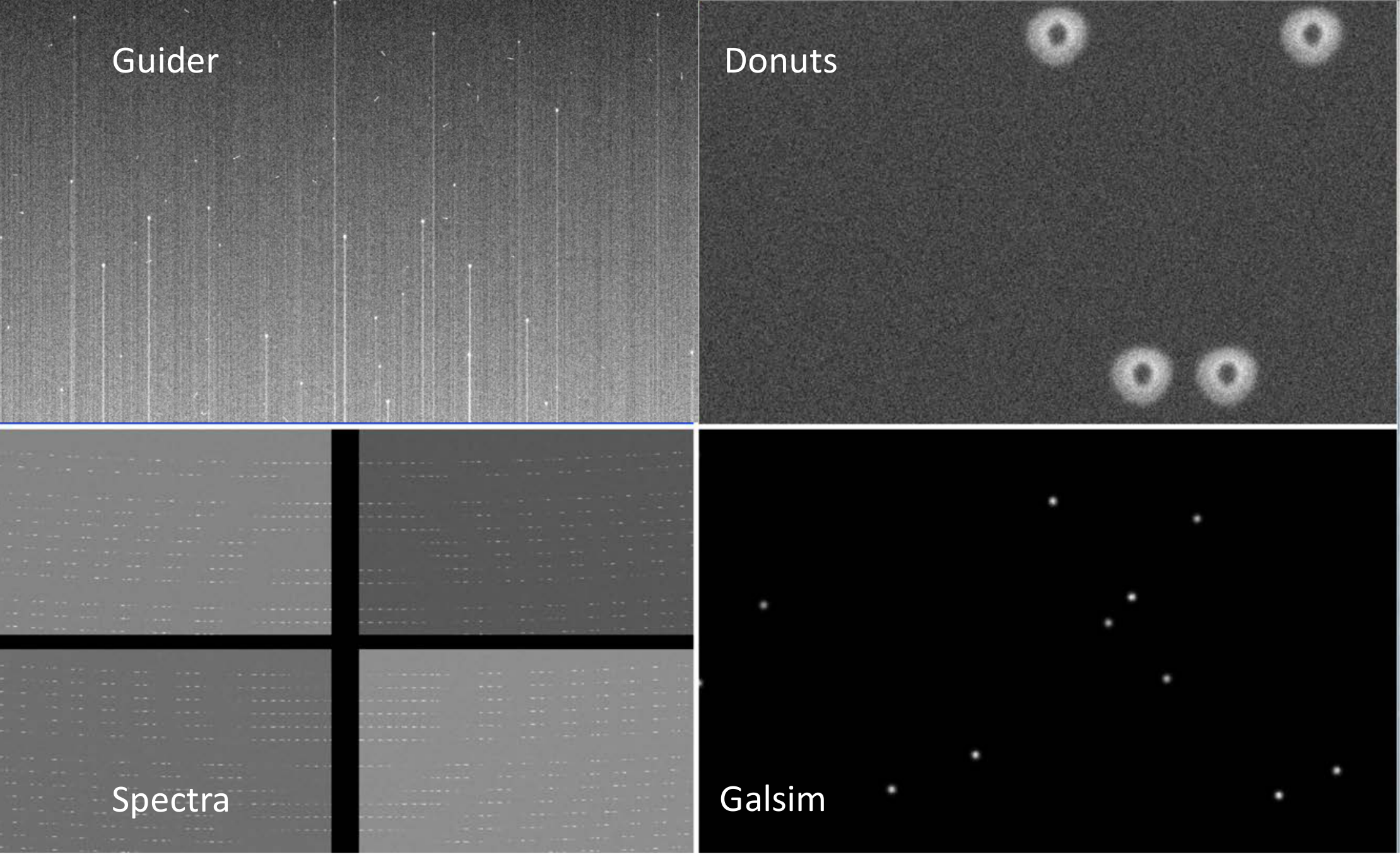}
\caption{Simulated images for the guider, focus and sky cameras and simulated spectra for the CCDs used for ICS full system tests.}
\label{fig:simulations}
\end{figure}

Using the DESI computer farm and observer consoles at the observatory, we are now performing mock observing tests of the entire ICS. Since the DESI instrument hardware is not yet available we developed simulated data sources - sample images are shown in Figure~\ref{fig:simulations} - and software emulators for every device controller and hardware component. 

These tests will also assess the online/mountain-top data quality tools provided by the ICS.
Continuous monitoring of both the hardware and the data quality is necessary to control systematic uncertainties at the level required to achieve the science goals of the project and to allow continuous, error-free operation of the DESI instrument.  
\begin{figure}[htb]
\centering
\includegraphics[width=0.8\textwidth]{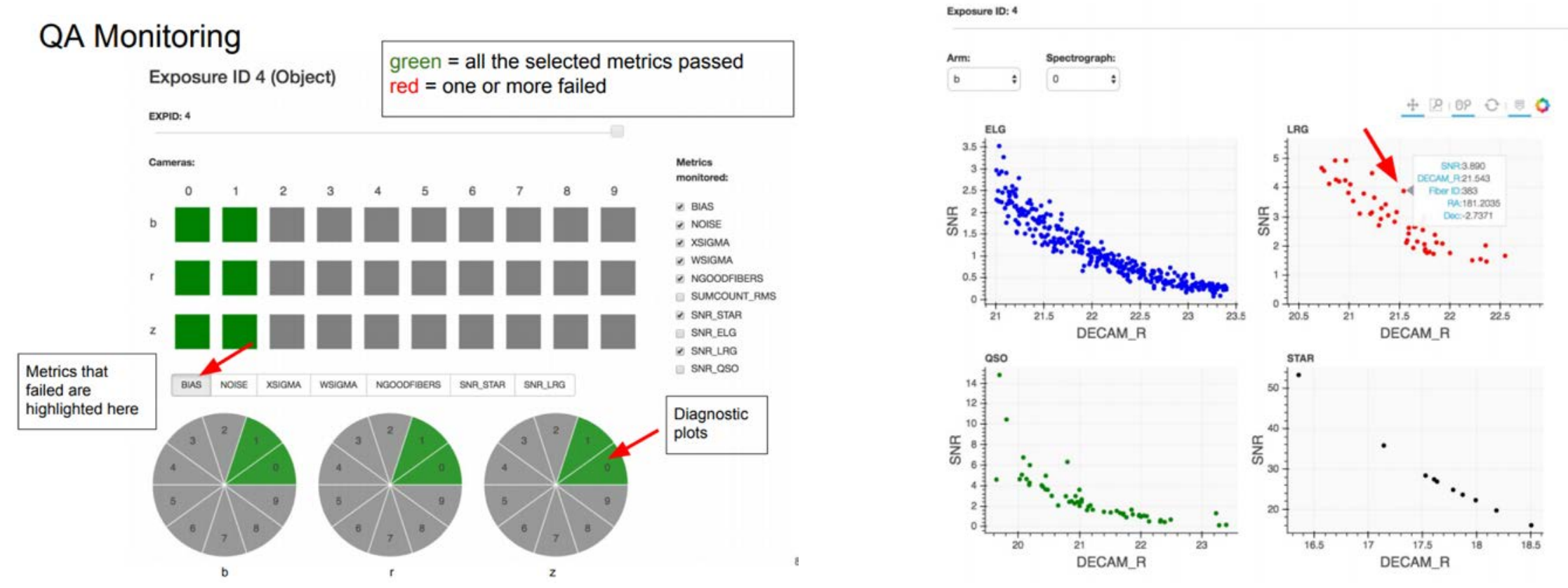}
\caption{Sample output of the QuickLook quality assessment tool. During observing, QuickLook analyzes every exposure providing near real-time feedback to the observer.}
\label{fig:quicklook}
\end{figure}
During an observing night, feedback to the observers is provided via the ICS GUIs such as the Observer Console and the QuickLook quality assurance tool that processes every exposure and provides near real-time feedback on the data quality to gauge survey progress and to ensure that the instrument is performing as expected (Figure~\ref{fig:quicklook}).
Building on experience gained with BOSS and DES, the QuickLook pipeline is a simplified version of a full offline reduction pipeline, replacing the most expensive computational steps with simpler
(and in some cases, more robust) algorithms.
QuickLook determines sky levels and calculates S/N estimates per exposure as a function of
wavelength and object magnitude.  This allows a robust
assessment of data quality and provides a cross check of the exposure length and
tile completeness as determined by the guider data and exposure time calculator.

\begin{figure}[htb]
\centering
\includegraphics[width=0.8\textwidth]{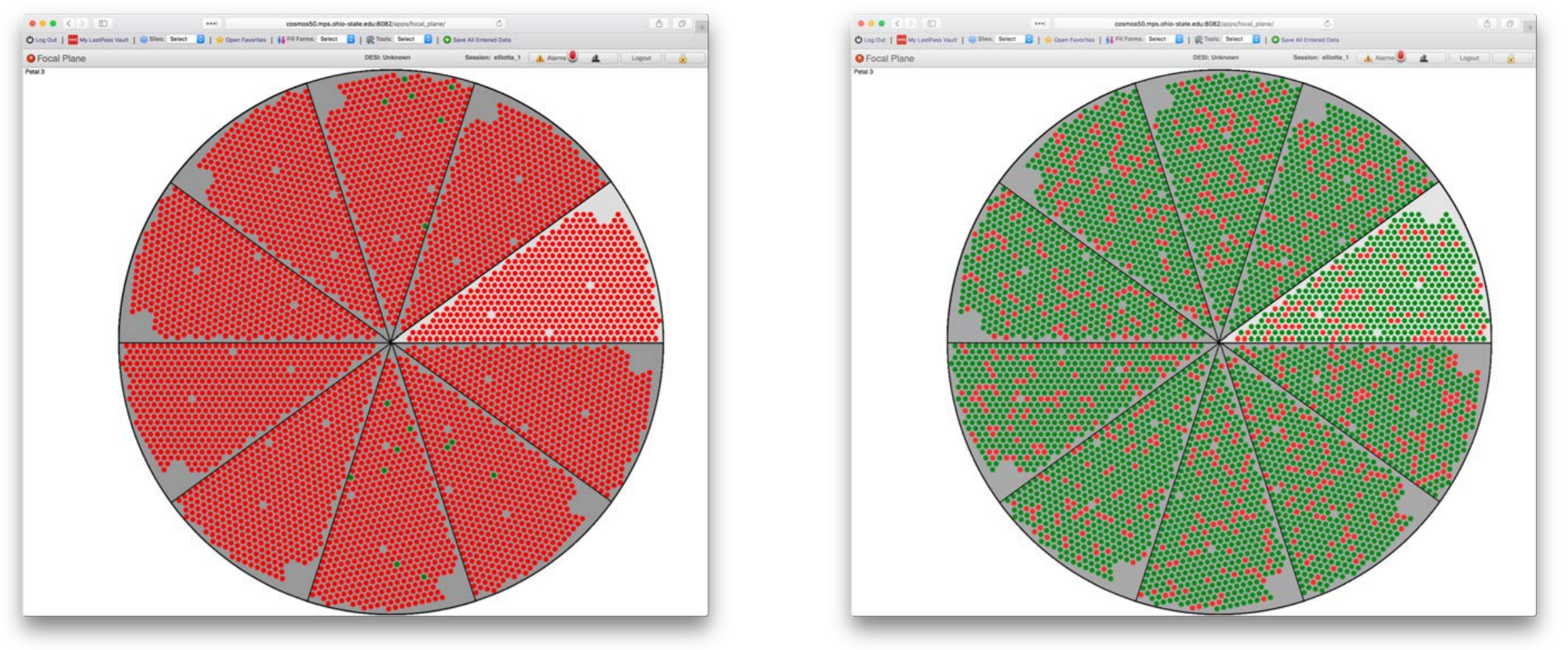}
\caption{An example of a DESI user interface showing the status of the focal plane in a browser window. The display changes dynamically during the fiber positioning loop. For each positioner, red indicates out of position and green indicates on-target.}
\label{fig:fpagui}
\end{figure}
The ICS provides the set of user interfaces required to operate the DESI instrument.
This set of graphical user interfaces includes elements for system and exposure control, alarm displays and telemetry monitors. The DESI observer console GUI acts as the primary observer interface for day to day operation. Combined, they present the user with the most commonly used information and everything that is needed to operate the DESI instrument. DESI GUIs are web-based and implemented in JavaScript and HTML/CSS using  standards such as HTML5 and web-sockets to provide the responsiveness expected from a modern system. Using this web-based approach provides many desirable features such as platform independence, remote access, a large number of 3rd party tools, and a certain level of security. 
The ICS GUIs are developed using the SproutCore HTML5 application framework.
As an example, Figure~\ref{fig:fpagui} shows the a display of the DESI focal plane. For each of the 5,000 fiber positioners a green circle indicates on-target whereas positioners that have not yet converged are shown in red. The display updates dynamically during the fiber positioner loop providing real-time feedback to the observers.

The goal of these mock observing tests is to demonstrate operational functionality by the end of June 2018 when the project transitions to installation and commissioning.

\section{Summary and Outlook}

 In this paper we reviewed the design and current status of DESI instrument control system. DESI is  a new powerful multi-object fiber spectrograph that will be installed on the Mayall 4m telescope at KNPO in 2019. 
 A fully operational ICS is required before that date to support functional validation of DESI components arriving at the observatory, installation and on-sky commissioning. To ensure in time completion of the ICS project, we devised a series of increasingly complex tests that were designed to validate the ICS infrastructure software, architecture and core components.  Details of these tests were presented in this paper. As we are approaching the end of the construction project we are now performing full system tests using the DESI computer hardware already installed in the Mayall dome and simulated data sources and devices controllers. The ICS will be ready for the start of DESI commissioning.

\section{Acknowledgements}

This research is supported by the Director, Office of Science, Office of High Energy Physics of the U.S. Department of Energy under Contract No. 
DE$-$AC02$-$05CH1123, and by the National Energy Research Scientific Computing Center, a DOE Office of Science User Facility under the same 
contract; additional support for DESI is provided by the U.S. National Science Foundation, Division of Astronomical Sciences under Contract No. 
AST-0950945 to the National Optical Astronomy Observatory; the Science and Technologies Facilities Council of the United Kingdom; the Gordon 
and Betty Moore Foundation; the Heising-Simons Foundation; the National Council of Science and Technology of Mexico, and by the DESI 
Member Institutions.  The authors are honored to be permitted to conduct astronomical research on Iolkam Du'’ag (Kitt Peak), a mountain with 
particular significance to the Tohono O'’odham Nation. 
% References
\bibliography{bibliography}
\bibliographystyle{spiebib} % makes bibtex use spiebib.bst
\end{document}